\documentclass[%
 aip,
 amsmath,amssymb,
 reprint,%
]{revtex4-1}

\usepackage{graphicx}% Include figure files
\usepackage{dcolumn}% Align table columns on decimal point
\usepackage{bm}% bold math
\usepackage[utf8]{inputenc}
\usepackage[T1]{fontenc}
\usepackage{mathptmx}
\usepackage{etoolbox}

\makeatletter
\def\@email#1#2{%
 \endgroup
 \patchcmd{\titleblock@produce}
  {\frontmatter@RRAPformat}
  {\frontmatter@RRAPformat{\produce@RRAP{*#1\href{mailto:#2}{#2}}}\frontmatter@RRAPformat}
  {}{}
}%
\makeatother
\begin{document}

\preprint{AIP/123-QED}

\title{A buoyancy-drag model with a time-varying drag coefficient for evaluating bubble front penetration depth} %Title of paper



\author{Dongxue Liu}
%\thanks{}
\affiliation{Department of Plasma Physics and Fusion Engineering, University of Science and Technology of China, Hefei 230026, People’s Republic of China
}
\author{Tao Tao}
\affiliation{Department of Plasma Physics and Fusion Engineering, University of Science and Technology of China, Hefei 230026, People’s Republic of China
}
\author{Jun Li}
\affiliation{Department of Plasma Physics and Fusion Engineering, University of Science and Technology of China, Hefei 230026, People’s Republic of China
}
\affiliation{Collaborative Innovation Center of IFSA, Shanghai Jiao Tong University, Shanghai, 200240, People’s Republic of China}
\author{Qing Jia}
\affiliation{Department of Plasma Physics and Fusion Engineering, University of Science and Technology of China, Hefei 230026, People’s Republic of China
}
\author{Rui Yan}
\affiliation{Collaborative Innovation Center of IFSA, Shanghai Jiao Tong University, Shanghai, 200240, People’s Republic of China}
\affiliation{Department of Modern Mechanics, University of Science and Technology of China, Hefei 230027, People’s Republic of China}
\author{Jian Zheng}
\affiliation{Department of Plasma Physics and Fusion Engineering, University of Science and Technology of China, Hefei 230026, People’s Republic of China
}
\affiliation{Collaborative Innovation Center of IFSA, Shanghai Jiao Tong University, Shanghai, 200240, People’s Republic of China}
\email{jzheng@ustc.edu.cn}


\date{\today}

\begin{abstract}
To evaluate and control bubble front penetration depth $h_{B}$ induced by ablative Rayleigh-Taylor instability (ARTI) from a weakly nonlinear phase to a self-similar phase, we first propose an improved buoyancy-drag (BD) model with a time-varying drag coefficient. The coefficient incorporates the influence of multiple physical mechanisms, including non-steady ablation, preheating, and other mechanisms during this phase. The model is validated through simulations under various conditions, demonstrating improved accuracy compared to the classical BD model and the self-similar growth. Furthermore, the model suggests controlling $h_{B}$ by suppressing the "most dangerous mode", which is influenced by initial perturbations and ablative acceleration history, thus offering novel insights for target manufacturing and pulse optimization near the ignition threshold.
\end{abstract}

%\pacs{jzheng@ustc.edu.cn}% insert suggested PACS numbers in braces on next line

\maketitle %\maketitle must follow title, authors, abstract and \pacs

\section{Intruduction}
%\label{}
Inertial confinement fusion (ICF) aims to achieve high-gain ignition\cite{abu2022lawson,abu2024achievement}, which remains a significant challenge primarily due to the influence of hydrodynamic instabilities\cite{smalyuk2019review,casner2021recent}. Among these instabilities, ablative Rayleigh-Taylor instability (ARTI) is particularly important, as it can evolve from a linear phase into a nonlinear phase during the acceleration phase. Nonlinear ARTI can lead to interpenetration or mixing between the hot, low-density ablator and the cold, dense ablator or fuel, for\text{min}g bubbles that penetrate the implosion shell. It is recognized that the appearance of bubbles is one of the main reasons degrading yield-over-clean (YOC) in early ignition experiments conducted during the National Ignition Campaign (NIC)\cite{haan2011point,lindl2014review,hurricane2014fuel} and has recently been implicated in target gain discrepancies\cite{divol2024thermonuclear,pak2023overview} among shots with similar targets and lasers. Previous studies have demonstrated that target defects and laser imprint\cite{goncharov2006early,liu2022mitigating} could evolve into a highly nonlinear phase\cite{smalyuk2005fourier,casner2015probing}, even a self-similar phase\cite{zhang2018self,zhang2020nonlinear,martinez2015evidence,sadot2005observation}. However, a unified growth model for evaluating and controlling multimode ARTI from the weak nonlinear phase to the self-similar phase in real space is still lacking.

The theories of ARTI in various phases can be summarized as follows. In the linear phase, a single mode grows exponentially until reaching an amplitude about 1/10 to 1/5 of its wavelength. The linear growth rate shapes like \(\gamma \cong \sqrt{A_t kg}-\beta kv_a\)\cite{takabe1985self,betti1996self}, where $v_a$ , $g$ , $k$ , $A_t$ and $\beta $ are the ablative velocity, acceleration, perturbation wave number, Atwood number, and a factor that depends on the heat conduction power index $\nu $, respectively. In the nonlinear phase, a single mode achieves a ter\text{min}al velocity, ${{U}_{B}}= C\sqrt{gA_t \lambda }$\cite{goncharov2002analytical}, where $C$ is a factor dependent on dimensions and $\lambda $ represents perturbation wavelength. For a full spectrum of perturbation modes, Hann\cite{haan1989onset} proposes that nonlinear behavior initiates when the cumulative amplitude of modes within a wave packet becomes comparable to a wavelength. The evolution of the wave packet could be approximated as a single mode, which has been validated experimentally\cite{sadot2005observation}. Furthermore, Hann’s weakly nonlinear theory for broadband ARTI\cite{haan1991weakly}, which incorporates $\gamma $ into a classical second-order expansion, demonstrates that ablation can effectively suppress the nonlinear growth rate. Since high-order expansions are inadequate for resolving the evolution of highly nonlinear ARTI, based on Hann’s work, Dimonte\cite{dimonte2004dependence} has proposed a method to derive the self-similar coefficient necessary for evaluating the bubble front penetration depth ${{h}_{B}}$ during the subsequent self-similar phase. Nevertheless, the calculated self-similar coefficient varies due to non-steady ablation and complex physics inherent to ICF, indicating that the system is transiting towards a self-similar phase. To evaluate ${{h}_{B}}$ in this phase, the classical buoyancy-drag (BD)\cite{baker1981heuristic,dimonte2000spanwise} model, which can be derived from Layzer’s potential flow model\cite{layzer1955instability}, provides a valuable framework. 

Originally proposed by Young\cite{hansom1990radiation} as a forcing equation for bubble dynamics, the BD model is adaptable due to several coefficients with physical connotations. In scenarios where $A_t =1$, Layzer’s model simplifies to a form: (mass + added mass) × acceleration = buoyancy – drag, detailed as follows\cite{layzer1955instability},
\[\left( 1+E \right)\frac{d{{V}_{B}}}{dt}=\left( 1+E \right)g-\frac{{{C}_{d}}}{R}V_{B}^{2},\]
where, ${{V}_{B}}=\frac{d{{h}_{B}}}{dt}$ represents the bubble velocity; $R$ relates to ${{h}_{B}}$ ; and an exponential term $E$ , applicable only in the linear phase with small amplitude, is omitted during the nonlinear phase. While both inertia and buoyancy are volumetric, drag is proportional to the cross-sectional area, incorporating a drag coefficient relevant to ${{C}_{d}}$. The added mass associated with bubble shapes arises from the movement of the penetrated fluid. Several extended models have been proposed, including modification of the transport process in the simulations to evaluate spikes at the interface of the hot spot\cite{rana2017mixing} and incorporation of an ablative stabilization term\cite{huntington2018ablative} to better capture the dynamics of bubbles influenced by radiative shocks. However, these models with constant drag coefficient often inadequately calculate ${{h}_{B}}$ from a weakly nonlinear phase to a self-similar phase, as this phase is influenced by diverse physical mechanisms\cite{li2022mitigation,li2024effect,zhang2022self,walsh2017self,walsh2023nonlinear}, such as non-steady ablation, non-local electron heat transport, hot electron preheating, and self-generated magnetic fields.

In this paper, to evaluate ${{h}_{B}}$ from a weakly nonlinear phase to a self-similar phase, we propose an improved BD model with a time-varying drag coefficient, which encompasses the influence of various physical mechanisms. We employ radiation-hydrodynamic simulations to validate the model in ICF. The model suggests controlling ${{h}_{B}}$ by suppressing the "most dangerous mode", which is influenced by initial perturbations and ablative acceleration history. This approach offers novel insights for target manufacturing and pulse optimization near the ignition threshold, particularly as the National Ignition Facility (NIF) seeks to enhance the \text{max}imum target gain through improved target quality\cite{pak2024observations}.

The paper is organized as follows. In Section II, we propose an improved BD model with a time-varying drag coefficient. Validations of the model and its further application in guiding the control of ${{h}_{B}}$ are presented in Section III. Finally, we draw our conclusions in Section IV.
\section{The improved buoyancy-drag model with a time-varying drag coefficient}
The classical BD model with a constant drag coefficient cannot accurately evaluate the bubble evolution from a weakly nonlinear phase to a self-similar phase. We propose an improved BD model with a time-varying drag coefficient, which involves the influence of multiple physical mechanisms through $\gamma $ and ${{U}_{B}}$ \cite{goncharov2002analytical} in this phase.

 \begin{figure}[h]
 \includegraphics[width=0.5\linewidth]{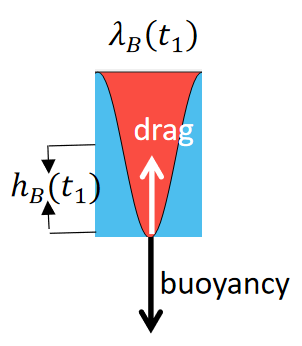}%
 \caption{\label{FIG. 1} Forcing of a bubble with the red indicating low-density fluid and the blue representing dense fluid. $\lambda_{B}$ represents bubble wavelength.}%
 \end{figure}
The classical BD model has been well applied to weakly nonlinear phase and its subsequent phases\cite{el2020numerical,youngs2020buoyancy}, particularly in describing the dynamics of bubble forcing, as illustrated in Fig.  \ref{1}. Various drag coefficients are utilized throughout these phases, typically expressed as follows\cite{dimonte2000spanwise}, 
\[D\sim {{C}_{d}}\frac{{{\rho }_{B}}}{{{\rho }_{S}}+{{\rho }_{B}}}\frac{{{h}_{B}}}{{{\lambda }_{B}}},\]
where $\rho_{B}$ and $\rho_{S}$  denote the bubble density and spike density. 
 
In the context of steady ablation, early experimental results have shown that the average ${{h}_{B}}/{{\lambda }_{B}}$ and ${{\rho }_{B}}/{{\rho }_{S}}$\cite{sadot2005observation} remains invariant, thus $D$ is approximately constant from the weakly nonlinear to self-similar phases. Consequently, we can use the specific self-similar behavior to derive the expression for $D$ within this steady-ablation scheme. The derivation of self-similar behavior, do\text{min}ated by bubble competition\cite{zhang2018self,zhang2020nonlinear,martinez2015evidence}, was based on established research: Hann’s theory\cite{haan1989onset} suggests representing a full mode spectrum with a do\text{min}ant mode; Fermi’s nonlinear transition\cite{layzer1955instability} theory indicates that a single-mode perturbation experiences exponential growth during the linear phase until it reaches ter\text{min}al velocity ${{U}_{B}}=C\sqrt{{{A}_{t}}g{{\lambda }_{B}}}$\cite{goncharov2002analytical}; and Birkhoff’s approach\cite{birkhoff1955taylor} obtains the self-similar evolutions of the do\text{min}ant bubbles by seeking ${{\lambda }_{B}}$ that \text{max}imizes ${{h}_{B}}$. The self-similar coefficient is expressed as follows,
\begin{equation}
\begin{split}
  {{\alpha }_{ab}}&=\frac{\Xi C\sqrt{\pi }}{4}{{\left( ln\left( {{U}_{B,0}}/\left( \left\langle {{h}_{k0}} \right\rangle {{\gamma }_{k0}} \right) \right)-1 \right)}^{-1}} \\ 
 & =\frac{\Xi C\sqrt{\pi }}{4}{{\left( ln\left( C\sqrt{\pi }/\left( {{k}_{B,0}}\left\langle {{h}_{k0}} \right\rangle \Xi  \right) \right)-1 \right)}^{-1}}, \label{1}   
\end{split}
\end{equation}		
where \[\Xi ={{\gamma }_{k0}}/\sqrt{{{A}_{t}}{{k}_{B0}}g}=1-\beta \sqrt{{{k}_{B,0}}/\left( {{A}_{t}}g \right)}{{v}_{a}}\] is the linear growth rate of ARTI normalized with that of classical RTI and $\left\langle {{h}_{k0}} \right\rangle $ represents the initial perturbations at the onset of acceleration. If $\left\langle {{h}_{k0}} \right\rangle $ is sufficiently small to remain in the linear phase initially, as indicated by the positive denominator in equation (1), this coefficient could be derived\cite{dimonte2004dependence} and validated by simulations\cite{zhang2018self}. To get the self-similar growth of $h_B$, 
\begin{equation}
{{h}_{B}}={{\alpha }_{ab}}{{A}_{t}}{{\left( \int{\sqrt{g}dt} \right)}^{2}},    \label{2}
\end{equation}	
from Youngs’s BD model\cite{hansom1990radiation}, 
\begin{equation}
\frac{d{{V}_{B}}}{dt}={{A}_{t}}g-D\frac{{{V}_{B}}\left| {{V}_{B}} \right|}{{{h}_{B}}},  \label{3} 
\end{equation}
we first derive the expression of the time-varying drag coefficient by substituting equation (\ref{2}) into (\ref{3}),
\begin{equation}
D=\frac{1}{4}\left( \frac{1}{{{\alpha }_{ab}}}-\frac{2s}{s+2}-2 \right),    \label{4}
\end{equation}
where $g\sim {{t}^{s}}$, $(s\ge 0)$, and $s=0$ and a constant $\Xi$ means steady ablation. Consequently, the improved BD model is articulated using equation (\ref{3}) with equation (\ref{1}) and (\ref{4}). If $D$ is calculated with a constant $g$ while ${{h}_{B}}$ evolves with a time-varying $g$, ${{h}_{B}}$ would be underestimated for $s>0$ due to a larger $D$.

In the context of non-steady ablation, various physical mechanisms, such as ablation, non-local electron heat transport, hot electron preheating, and self-generated magnetic fields, influence the growth of ARTI through the two factors of $\gamma \left( t \right)$ and ${{U}_{B}}\left( t \right)$\cite{betti2006bubble,walsh2023nonlinear}. Consequently, the history of ablative acceleration and related physical processes can be incorporated through variations in $\gamma \left( t \right)$ and ${{U}_{B}}\left( t \right)$ as indicated in equation (\ref{1}). 

 \begin{figure}[h]
 \includegraphics[width=1\linewidth]{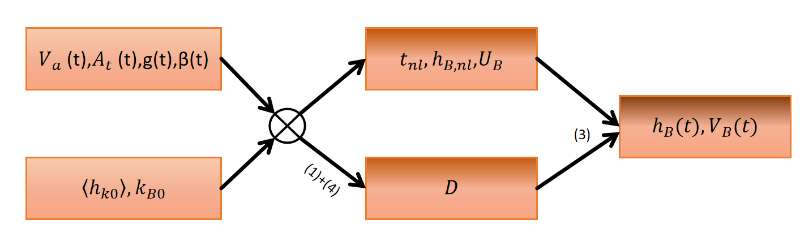}%
 \caption{\label{FIG. 2} The calculation process of our model. Here, the subscript ‘nl’ represents the onset of nonlinearity.}%
 \end{figure}
As illustrated in Fig.  \ref{FIG. 2}, to evaluate ${{h}_{B}}$, deter\text{min}ing the initial do\text{min}ant perturbation is essential for calculating $D$ and initiating equation (\ref{3}). Given that the spectrum of initial multimode perturbations in ICF may consist of several wave packets, based on Hann’s study, we conceptualize these spectrums as a combination of distinct wave packets, each centered around its respective do\text{min}ant wavelength. The wave number of the do\text{min}ant perturbation can be extracted from the simulated areal density distribution, \[{{k}_{B0}}=\int{\rho _{k0}^{2}{{Y}^{2}}dk/\left( \sum{\rho _{k0}^{2}{{Y}^{2}}} \right)}.\] However, $\left\langle {{h}_{k0}} \right\rangle $ is notably small, ranging from ${{10}^{-8}}$ cm to ${{10}^{-5}}$ cm \cite{dimonte2004dependence}, necessitating alternative methods for its deter\text{min}ation. The perturbation velocity at the onset of acceleration is taken as ${{v}_{\text{k0}}}=U-\left\langle U \right\rangle $\cite{peebles2019direct} and the initial perturbation amplitude is expressed as $\left\langle {{h}_{k0}} \right\rangle \approx {{v}_{\text{k0}}}/\gamma $, where $U$ represents local fluid velocity and $\left\langle U \right\rangle $ denotes mass weight mean fluid velocity. With these initial perturbations, as well as data from one-dimensional simulations, especially ${{v}_{a}}$(t), ${{A}_{t}}$(t), $g$(t) and $\beta$(t), the calculated ${{h}_{B}}$ during nonlinear phase can be validated through following simulations.
\section{Validation and application of the model via simulations}
\subsection{Simulation setup}
In this study, we employ the two-dimensional Eulerian radiation-hydrodynamic code FLASH\cite{fryxell2000flash} to validate the improved BD model represented by equation (\ref{3}), which can guide the control of ${{h}_{B}}$ in ICF. The key simulation setup is as follows. The lengths of the simulation domain in the $X$ and $Y$ directions are denoted as ${L}_{x}=$[ -100 $\mu$m , 100 $\mu$m ] and ${{L}_{y}}=$[ -400 $\mu$m, 2000$\mu$m] , respectively, with a spatial resolution of 0.52 $\mu$m. The laser pulse $^\text{a)}$, vertically irradiating the planar target, is a square pulse with a rise time of 0.1 ns and a peak intensity of 25 TW/cm$^2$. The CH planar target is set with a density of 1 g/cc and a thickness of 90 $\mu$m . For the seeding sources of multi-mode velocity perturbations adjacent to the surface, we define them as ${{V}_{p}}\left( x \right)=\sum {{V}_{pk}}cos\left( m{{k}_{L}}x+{{\psi }_{k0}} \right)$, ${{k}_{L}}=2\pi /{{L}_{x}}$, where $m$ is an integer ranging from 4 to 10, ${{\psi }_{k0}}$ is a random phase uniformly distributed between zero and one, ${{V}_{p}}k={{V}_{p}}k0{{e}^{\left( -m{{k}_{L}}|y-{{y}_{0}}| \right)}}$, ${{V}_{pk0}}=F{{\left( m{{k}_{L}} \right)}^{-2}}$, and $F$ is a constant. Table I displays the information of non-ideal simulations, which encompass the process of laser energy deposition, electron heat conduction, and heat exchange between electrons and ions, while excluding radiation transport effects. For case 1 at t = 3.4 ns, ${{V}_{p}}\left( x \right)$ has been loaded onto the target and a shock wave propagates through the rear interface of the target at $Y=0$ $\mu$m , leading to the rapid growth of perturbations. 
\begin{table}[h]
    \centering
    \begin{tabular}{ccccc}
    \hline
         case&  pulse&   $ \left\langle {{h}_{k0}} \right\rangle \approx {{v}_{k0}}/\gamma $ 
&   $ {{k}_{B0}} $ ( $ 1{{0}^{3}}c{{m}^{-1}} $ )& independent variable\\
\cline{1-5}
         1&  $^\text{a)}$&  2.33&  2.43& \\
         \cline{1-5}
         2&  $^\text{b)}$&  2.33&  2.43& 1→2:pulse$^\text{a)}$→pulse$^\text{b)}$\\
         \cline{1-5}
         3&  $^\text{c)}$&  10&  2.94& \\
         \cline{1-5}
         4&  $^\text{d)}$&  10&  2.94& 3→4:pulse$^\text{c)}$→pulse$^\text{d)}$\\
         \cline{1-5}
         5&  $^\text{a)}$&  6.62&  2.33& 1→5:random\\
         \cline{1-5}
         6&  $^\text{a)}$&  10.4&  2.26& 1→6:deliberate\\
         \cline{1-5}
         7&  $^\text{a)}$&  23&  2.32& 1→7:deliberate\\
         \cline{1-5}
         8&  $^\text{a)}$&  16.00&  2.73& \\
         \cline{1-5}
         9&  $^\text{a)}$&  16.80&  4.95& 8→9:deliberate\\
         \cline{1-5}
 10& a)& 14.30& 2.17&9→10:deliberate
\\
\cline{1-5}
 11& c)& 10.00& 2.94&3→11: =0.06→ =0.04
\\
\cline{1-5}
 12& c)& 10.00& 2.94&3→12:  =0.06→ =0.1

\\
\hline
    \end{tabular}
    \caption{The information of simulation cases with perturbations, where ‘deliberate’ represents modifying the initial perturbations with $F$ and $m$ , while ‘random’ denotes modifying the initial perturbation with random phase and $f$  represents the flux-limited value.}
    \label{表格 5.1}
\end{table}
\subsection{Validation of the model}
\begin{figure}[h]
 \includegraphics[width=1\linewidth]{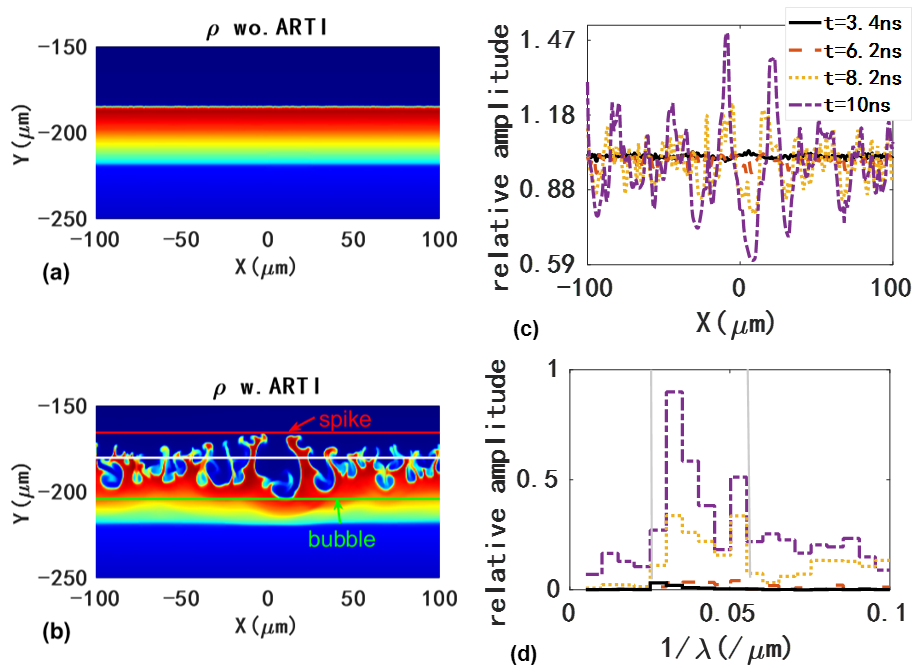}%
 \caption{\label{FIG. 3} Distribution of density $\rho $ for cases with (b) and without (a) velocity perturbations at 10 ns , where the red, white and green lines represent the spike front, the boundary between the spikes and the bubbles and the bubble front, respectively. The spatial distribution (c) and Fourier spectrum (d) of $\rho $ in $X$ after summed in $Y$. The black solid lines correspond to the initial perturbations with their spectrums bounded by two gray lines in (d). The colors and the line styles are the same in (c) and (d). The dominant perturbation wavelengths at t = 10 ns are identical as those of the initial, demonstrating that bubble competition drives the evolution of the bubble front.}%
 \end{figure}

We conduct detailed validation of the model through simulations, focusing on the bubble competition mechanism, ablative stabilization effects, applicability in non-steady ablation, and the sensitivity of $D$ to initial perturbations and ablative acceleration history. In comparison to Fig. \ref{FIG. 3}(a), the multi-mode velocity perturbations in Fig. \ref{FIG. 3}(b) lead to significant deformation of the planar target, exhibiting nonlinear characteristics at t = 10 ns. The low-density fluid penetrates the dense fluid, resulting in the formation of bubbles, whereas the dense fluid penetrates the low-density fluid, for\text{min}g spikes. The interface between the bubbles and the spikes is defined as the outer edge of the accelerated shell, represented by the average position with $1/e$ of the peak density near the coronal region. The bubble front is expressed as $Y=\text{min}\left( \text{min}\left( {{\nabla }_{Y}}\rho (X,Y) \right) \right)$, and the spike front is denoted by $Y=\text{max}\left( \text{max}\left( {{\nabla }_{Y}}{{T}_{e}}(X,Y) \right) \right)$, where ${{T}_{e}}$ represents the electron temperature.

Fig. \ref{FIG. 3}(c) and \ref{FIG. 3}(d) illustrate the evolution of bubbles in both real and spectral space to validate the bubble competition. The black solid lines represent the initial perturbations, with their spectrums bounded by two gray lines in Fig. \ref{FIG. 3}(d), at the onset of acceleration (t = 3.4 ns). From t = 3.6 ns to t = 6.2 ns, the initial perturbations with smaller wavelengths exhibit faster growth. Upon entering the nonlinear phase at t = 6.2 ns, the initial perturbations with larger wavelengths begin to do\text{min}ate due to competitive interactions among bubbles, leading to the phenomenon where one bubble expands while its neighbors shrink. Although perturbations with wavelengths exceeding the initial perturbation wavelengths emerge due to bubble merging, bubble competition continues to do\text{min}ate the evolution, as the amplitude of perturbations, outside the range bounded by two grey lines, remains smaller than that of the initial perturbations. Consequently, relating drag coefficient $D$ to the initial perturbations is justified.

 \begin{figure}[h]
 \includegraphics[width=1\linewidth]{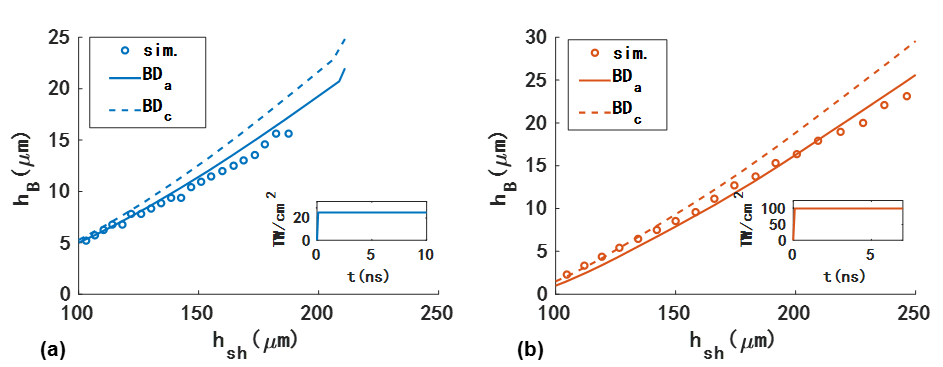}%
 \caption{\label{FIG. 4} The relationship between ${{h}_{B}}$ and the distance, ${{h}_{sh}}$ , traveled by the ablation front that is also the interface between the bubble and the spike for (a) : case 1 and (b) : case 3. The symbols are obtained from simulations, and the dashed and solid lines are calculated  using equation (\ref{3}) without and with ablation, respectively. The insets are laser pulses used in simulations.}%
 \end{figure}
In the context of steady ablation, we validate the ablative stabilization of equation (\ref{3}). ${{h}_{B}}$ calculated by equation (\ref{3}) without ablation is larger than that obtained from simulations, while the calculated ${{h}_{B}}$ from equation (\ref{3}) with ablation aligns well with the simulation results under various laser pulses in Fig.\ref{FIG. 4}(a) and (b). As the distance $h_{sh}$ traveled by the ablation front increases, the effects of accumulated ablative stabilization become more pronounced.

\begin{figure}[h]
 \includegraphics[width=1\linewidth]{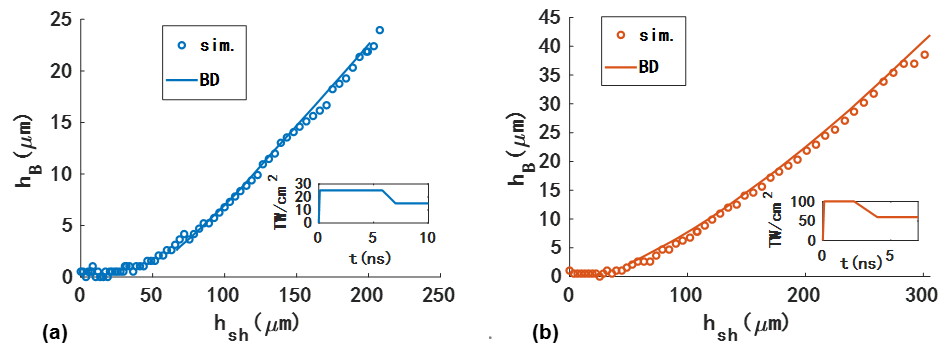}%
 \caption{\label{FIG. 5} The relationship between ${{h}_{B}}$ and ${{h}_{sh}}$ in non-steady situations for (a): case 2 and (d): case 4.}%
 \end{figure}

In the context of non-steady ablation, the calculated ${{h}_{B}}$ corresponds to the simulation results under two different pulses, as depicted in Fig. \ref{FIG. 5}(a) and (b). It is evident that equation (\ref{2}) does not provide better predictions than equation (\ref{3}) due to the non-linear relation between ${{h}_{B}}$ and ${{h}_{sh}}\sim {{\left( \int{\sqrt{g}}dt \right)}^{2}}$ . Additionally, resolution effects introduce errors ($\pm 1.04$ $\mu \text{m}$), leading to reduced confidence in ${{h}_{B}}$ before ${{h}_{sh}}\le 50$ $\mu \text{m}$.

\begin{figure}[h]
 \includegraphics[width=1\linewidth]{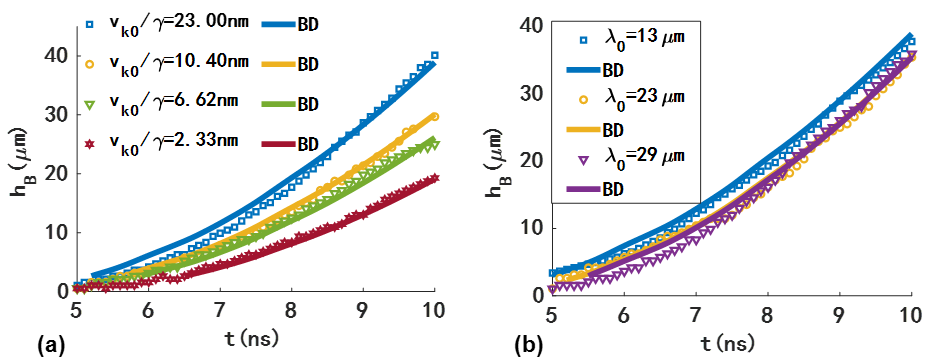}%
 \caption{\label{FIG. 6} The relationship between ${{h}_{B}}$ and $t$. (a) with different initial perturbation amplitudes ${{v}_{k0}}/\gamma $ for case 1, 5, 6 and 7; (b) with different initial perturbation wavelengths ${{\lambda }_{0}}=2\pi /{{k}_{B,0}}$ for case 8-10.}%
 \end{figure}

The time-varying drag coefficient $D$ is influenced by the initial perturbations and ablative acceleration history as seen in equation (\ref{1}) and (\ref{4}). For the same acceleration $g$ , ${{h}_{B}}$ increases as $D$ decreases due to a reduced drag force. Fig. \ref{FIG. 6}(a) illustrates the dependence on $\left\langle {{h}_{k0}} \right\rangle ={{v}_{k0}}/\gamma $ , where the onset of nonlinearity is indicated by the beginning of the solid fitting lines. Cases with larger $\left\langle {{h}_{k0}} \right\rangle $ transition into the nonlinear phase earlier and achieve greater ${{h}_{B}}$ simultaneously. The impact of $\left\langle {{h}_{k0}} \right\rangle $ becomes increasingly important for longer acceleration times\cite{casner2015probing}, during which bubbles can progress through several generations. This underscores the necessity for stringent  requirements in $\left\langle {{h}_{k0}} \right\rangle $ for high compression ignition schemes.
 
Fig. \ref{FIG. 6}(b) presents the dependence on the perturbation wavelengths $\lambda _0$ , which remain unstabilized. The cases with shorter $\lambda _0$ exhibit larger ${{h}_{B}}$ because they enter nonlinear phase earlier. However, the sensitivity diminishes for cases with wavelength differences of less than 6 $\mu$m under this simulation setup. This indicates that effective control of ${{h}_{B}}$ through perturbation wavelength requires exceeding a threshold value in wavelength differences. 

\begin{figure}[h]
 \includegraphics[width=0.6\linewidth]{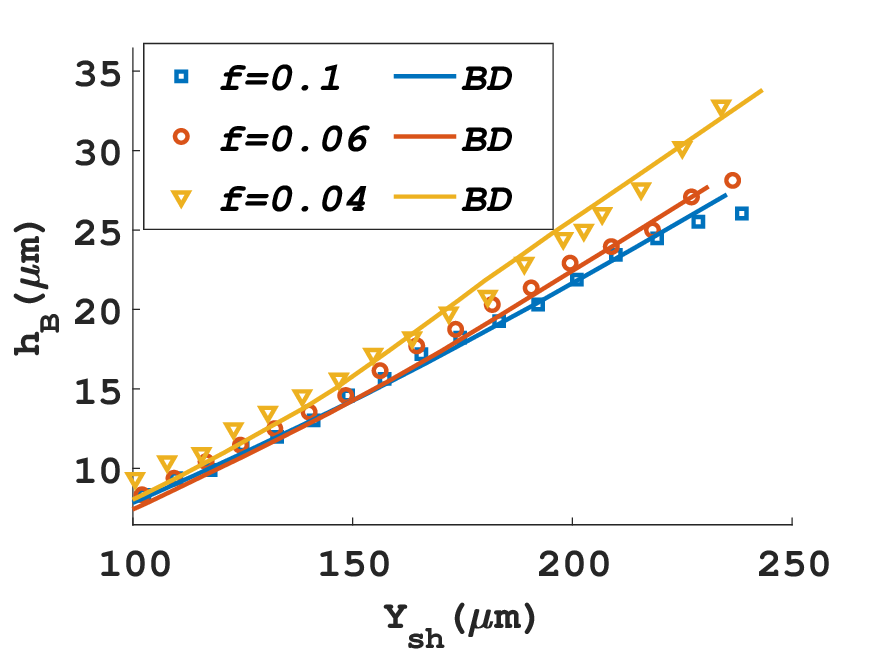}%
 \caption{\label{FIG. 7} ${{h}_{B}}$ varying with ${{Y}_{sh}}$ for case 12, 3 and 11. }%
 \end{figure}

To access the sensitivities of $D$ to ablative acceleration history, we modify the magnitude of local heat fluxes near the maximum temperature gradient by adjusting $f$ . Fig. \ref{FIG. 7} illustrates that until ${{h}_{sh}}>150$ $\mu \text{m}$, ${{h}_{B}}$ with different $f$ gradually diverges. The calculated ${{h}_{B}}$ aligns with the simulation results, increasing as $f$ decreases. If the drag force remains constant, $g$ increasing with $f$ leads to increasing perturbations, which contrasts with the trend shown in Fig. \ref{FIG. 7}. Thus, the elongation of bubbles at lower $f$ can be contributed to the decreasing drag force with a smaller $D$.
\subsection{Application in controlling bubble front penetration depth}
Although the model has been validated only in two-dimensional simulations, the various bubble growth observed in three-dimensional simulations can be incorporated in $D$ through ${{U}_{B}}$, which varies with distributions of vorticity and has different $C$ across different dimensions. Therefore, this model could be suitable for describing the evolution of bubbles in real situations. Following the validation of equation (\ref{3}), we will further explore strategies for controlling ${{h}_{B}}$. To illustrate this control strategy, we will consider self-similar growth as an example.

\begin{figure}[h]
 \includegraphics[width=0.6\linewidth]{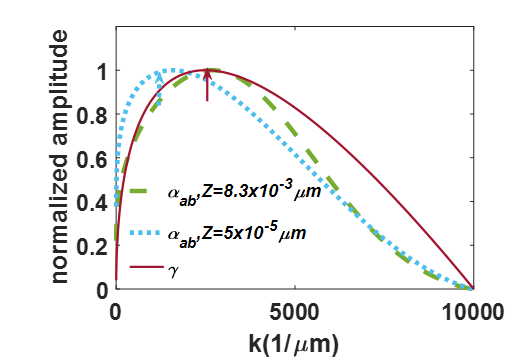}%
 \caption{\label{FIG. 8} Variation of ${{\alpha }_{ab}}$ and $\gamma $ with $k$ . The arrows represent the “most dangerous mode” . The initial value is taken as follows : $\beta \sqrt{1/\left( {{A}_{t}}g \right)}{{v}_{a}}={{10}^{-2}}\mu {\text{m}^{1/2}}$, $Z=\left\langle {{h}_{k0}} \right\rangle /\left( C\sqrt{\pi } \right).$}%
 \end{figure}

Equation (\ref{2}) indicates that maintaining ${{\alpha }_{ab}}$ below its maximum value can effectively control ${{h}_{B}}$. In the classical self-similar regime, which is dominated by bubble competition, the self-similar coefficient lacks an extremum due to $\frac{\partial {{\alpha }_{b}}}{\partial {{k}_{B}}}<0$\cite{dimonte2004dependence}. However, $\frac{\partial {{\alpha }_{ab}}}{\partial {{k}_{B}}}=0$ exists within the context of ARTI, leading to the establishment of the “most dangerous mode” ${{k}_{\alpha }}$ characterized by the maximum ${{\alpha }_{ab}}$, as illustrated by the blue arrow in Fig. \ref{FIG. 8}. Different from the “classical most dangerous mode” ${{k}_{\gamma }}$, determined by the maximum $\gamma $ , ${{k}_{\alpha }}$ depends on both initial perturbations and ablative acceleration history. This result indicates that it may be more reasonable for pulse optimization with our improved model instead of time-integrated $\gamma $ \cite{tao2023laser,zheng2022optimizing} as value function.

In the context of target defects or laser imprint, modes that initially exist in the nonlinear phase are unlikely to persist as the dominant mode during the nonlinear phase, as they will not exhibit a self-similar manner. Instead, these modes primarily influence the vorticity, which in turn affects the ${{U}_{B}}$ of ${{\lambda }_{0}}$ smaller than 10 $\mu$m\cite{betti2006bubble}. Therefore, based on $\left\langle {{h}_{k0}} \right\rangle $ generated by the current quality of the target and beam, we can suppressing  ${{k}_{\alpha }}$ as shifting from case 8 to 9 shown in Fig. \ref{FIG. 5}(b). This approach may elucidate target gain differences among shots with similar targets and provide novel insights for target manufacturing and pulse optimization near ignition threshold.
\section{Conclusions}
Prior research has demonstrated that target defects and laser imprint could evolve into a highly nonlinear, or even self-similar phase. In ICF, which typically involves non-steady ablation and various physical mechanisms, we first propose an improved BD model with a time-varying drag coefficient for evaluating and guiding the control of ${{h}_{B}}$ in real space. The coefficient incorporates multiple physical mechanisms through $\gamma$ and ${{U}_{B}}$. The model is validated through simulations under various conditions, demonstrating improved accuracy compared to the classical BD model and the self-similar growth. Furthermore, the model suggests controlling ${{h}_{B}}$ by suppressing the "most dangerous mode", which is influenced by initial perturbations and ablative acceleration history. This approach may elucidate the differences in target gains among shots with similar targets and lasers and provide novel insights for target manufacturing and pulse optimization near the ignition threshold.
\begin{acknowledgments}
This work is supported by the supported by the National Key R and D Projects (Grant No. 2023YFA1608402), Strategic Priority Research Program of the Chinese Academy of Sciences (Grant No. XDA25010200), and Nature Science Foundation of China (Grant No. 12375242).
\end{acknowledgments}
\section*{Data Availability Statement}
Data available on request from the authors.
% If in two-column mode, this environment will change to single-column format so that long equations can be displayed. 
%\begin{widetext}
%$$\mbox{put long equation here}$$
%\end{widetext}

% Figures should be put into the text as floats. 
% Use the graphics or graphicx packages (distributed with LaTeX2e).
% Put the label that you will use with \ref{} command in the braces of the \label{} command.
%
% \begin{figure}
% \end{figure}

% \begin{table}
% \end{tabular}
% \end{table}

%\begin{acknowledgments}
% Put your acknowledgments here.
%\end{acknowledgments}

\nocite{*}
\bibliography{main}

\end{document}